\documentclass[aps,pra,reprint,10pt,a4paper]{revtex4-1}
\usepackage[utf8]{inputenc}

\usepackage[sc,osf]{mathpazo}
\usepackage[T1]{fontenc}
\usepackage{amsfonts}
\usepackage{amssymb}
\usepackage{amsmath}
\usepackage{amsthm}
\usepackage{graphics}
\usepackage{graphicx}
\usepackage{braket}
\usepackage[normalem]{ulem}
\usepackage[colorlinks=true,linkcolor=blue,
urlcolor=blue,citecolor=blue]{hyperref}

\begin{document}
\title{Dimensional enhancements in a quantum battery with imperfections }

\author{Srijon Ghosh and Aditi Sen(De)}

\affiliation{Harish-Chandra Research Institute and HBNI, A CI of Homi Bhabha National Institute, Chhatnag Road, Jhunsi, Allahabad - 211019, India}

\begin{abstract}
We show that the average power output of a quantum battery based on a quantum interacting spin model, charged via a local magnetic field, can be enhanced with the increase of spin quantum number, thereby exhibiting dimensional advantage in quantum batteries. In particular, we demonstrate such increment in the power output when the initial state of the battery is prepared as the ground or canonical equilibrium state of the spin-$j$ $XY$ model and the bilinear-biquadratic spin-$j$ Heisenberg chain (BBH) in presence of the transverse magnetic field and  a weak value of interaction strength between the spins in the former model. Interestingly, we observe that in the case of the $XY$ model,  a trade-off relation exists  between the range of interactions in which the power increases and the dimension while for the BBH model, the improvements  depend on the phase in which the initial state is prepared.   Moreover, we exhibit that such dimensional advantages persist even when the battery-Hamiltonian has some defects or when the initial battery-state is prepared at finite temperature.

\end{abstract}

\maketitle

\section{Introduction}

Quantum computational  and communication devices are typically designed using two-level systems.   
It is also believed that a spin system with a large spin quantum number representing  a higher dimensional system looses its quantum nature, thereby showing a classical feature. On the contrary, it was shown that singlet state of arbitrary spin-\(j\) can violate Bell inequality maximally even when \(j\) is arbitrarily large \cite{gisinBell, gargBell}. 
Over last two decades, it was also demonstrated that higher-dimensional quantum systems, qudits,  can deliver  advantages over two-level systems in quantum protocols ranging from quantum computation, topological codes, quantum purification to quantum communication including quantum key distribution having improved key rate  \cite{sanders02, QKDqudits, QKDqudit2, QKDqudit1,  QCqudits, Bellqudits,  topologicalqudit, Maximalqudits,  Durpurify, highdimensionalteleportation, sappy} (see review on quantum technologies with qudits \cite{quditQtech}). For example, 
it was recently shown that the performance of a quantum switch, a device in which depending on the control qubit, operations are performed on the target qubit, can be enhanced by using qudits \cite{switchqudit}. Moreover, higher dimensional systems using physical substrates like photons \cite{photonexp}, ion traps \cite{ionqudits}, superconducting circuits \cite{supercondqudits}, nitrogen-vacancy centre \cite{NVcentrequdits} are prepared  in laboratories to exhibit  quantum information processing tasks.

From a different perspective, quantum spin models with
higher spin values have also attracted lots of attention.  One of the main reasons for such extensive investigations is  that the characteristics of the half-integer spin system can be completely different than that of the integer-spin models. Specifically, Haldane's conjecture  states that antiferromagnetic spin chains having  half-integer  and integer spins possess  contrasting properties in terms of excitation spectrum \cite{Haldane, Haldane1, Schulz, Parkinson'85, spin1, AkltPRL, aklt,  spin1XXZ}. Moreover, nonmagnetic phases like  Haldane, dimerized, nematic  phases are reported in spin models with arbitrary large spins compared to  a spin-$\frac{1}{2}$ chain.

All these results indicate that the quantum devices based on the quantum spin-$j$ model may lead to a better efficiency than that of the quantum spin-$\frac{1}{2}$ model. In this work, we will demonstrate that it is indeed the case by considering an energy  storage device, quantum battery \cite{Alicki, Batteryreview, AcinPRL, NJPSai,  KavanPRL, PoliniPRL, Modispinchain, PoliniPRL19,   OpenBattery1, adiabatic, highdim, ASD, OpenBattery2, albaprb, PoliniPRL19, marcoprl2, marcoprb1, marcoprb2, marcospringer, quach2020}. A ground state or canonical equilibrium state of an  interacting Hamiltonian is the initial state of the battery \cite{Modispinchain, ASD} while the energy is stored (extracted) in  (from) the battery by applying a local magnetic field. 
In recent years, a considerable amount of works have been carried out to understand the performance of the quantum battery in presence of decoherence, impurities, etc.,  and the property of the system which leads to quantum advantage. However,  all the investigations are restricted to multiqubit systems except for very few recent works on three-level systems \cite{adiabatic, highdim}. 


In this paper, we  consider a one-dimensional anisotropic transverse $XY$ quantum spin model \cite{spin1, piersPRB, XYJMM}  and bilinear-biquadratic Heisenberg (BBH) spin chain with arbitrary large spins \cite{Lai74, Takhta82, Suth75, Babu82, Fath91, Fath93, Buchta05, spin1:spin2} as a quantum battery. In particular, the system consists of $N$ spin -$j$ particles and  its ground state is taken  as the initial state which  evolves via a local charging field by  unitary evolution. We calculate the maximum extractable power from the battery by varying $j$ and find in both the models that power output is enhanced with the increase  of spin quantum number. However, the beneficiary character of quantum batteries in higher dimension is not ubiquitous. Specifically, we show both analytically and numerically that the improvement with high $j$ in terms of power can only be observed when the interaction strength is weak for the quantum $XY$ model and with low anisotropy. On the other hand, for the bilinear-biquadratic Heisenberg model, we again report   the dimensional advantage  although the enhancement depends on the phase of the initial state. 

From an experimental point of view, such a scenario  described above is ideal. We now introduce imperfections in two ways -- (1) finite-temperature state is considered as the initial state; and (2) impurity is present in the battery-Hamiltonian which is naturally arise during the preparation of the system. Surprisingly, we observe that for the $XX$ model, although the energy stored  (extracted) in (from) the  battery increases with the increase of dimension for the low-temperature regime, when the temperature in the thermal state is high, the opposite picture develops -- spin models with low spin quantum number exhibits higher power output compared to that of the higher dimensional battery. In presence of  defects  in the interaction strength which are site-dependent and chosen randomly from Gaussian distribution, we find that dimensional enhancement persists in the quenched average power  both for the $XX$ as well as bilinear-biquadratic models in presence of weak disorder strength quantified by the standard deviation of the  distribution.


The paper is presented as follows.  After describing the prerequisites at the beginning of Sec.  \ref{sec:enhance}, we analytically show in Sec. \ref{analytics} that battery built by using spin-$1$ $XX$ chain  results more output power than that  obtained via spin-$\frac{1}{2}$ $XX$ chain and then we establish the dimensional advantage for several values of $j$. The next section (in Sec. \ref{haldane}) reveals that the performance of the battery  depends both on the dimension and the phase of an isotropic bilinear-biquadratic Hamiltonian.  Sec. \ref{sec:imperfect} deals with the scenarios when the initial state is prepared at finite temperature or the battery-Hamiltonian has some randomness in the interaction strength. Finally, in Sec.  \ref{conclusion}, we summarize the results.

\section{Enhanced power of the battery with the increase of spin quantum number: Illustration by transverse $XY$ model}
\label{sec:enhance}

Quantum battery is modelled as a finite number of  quantum mechanical interacting systems in $d$-dimension, governed by a Hamiltonian, $H_{B}^j$, with $j$ being the spin quantum number, indicating the dimension of the battery. To charge the system, a local magnetic field, governed by the Hamiltonian, $H_{c}^j$, is applied  to each subsystem. 
The initial state, $\rho(0)$, of the battery, is taken to be the ground state or the canonical equilibrium state of the Hamiltonian $H_{B}^j$.

Our main motivation is to figure out the effects of higher dimensions on the energy storage and extraction processes  of the quantum battery. 
In a closed system, the total work output from the battery is defined as \cite{Alicki, Batteryreview} $W(t) = W_{final} - W_{initial} = \mbox{Tr}(H_{B}^j\rho(t)) -  \mbox{Tr}(H_{B}^j \rho(0))$, where the initial energy of the system is $\mbox{Tr}(H_{B}^j\rho(0))$ and \(\rho (t)\) represents the dynamical state of the battery  at time \(t\) which is obtained when the  local charger, $U(t)$  acts on the initial state $\rho(0)$, i.e., \(\rho (t) =   U(t)  \rho(0)  U(t)^{\dagger}\) with \(U (t) = e^{- i H_c^j t}\).  
Notice that the storing energy in the battery may not always be the same with the maximum energy that can be extracted from the battery, quantified via  ergotropy \cite{Alicki} (cf. \cite{PoliniPRL19, quach2020}). However, we will show that in our system, both the quantities coincide when the ground state of the battery acts as the initial state. 
The maximum average power output from the system, in our case,  reads as
 \(   P_{\max}= \max_{t} \frac{W(t)}{t}\).

Our main aim is to obtain a comparative study in the efficiency of the battery in terms of  $P_{max}$ with increasing the dimension of the system. To make all the models in arbitrary dimension in the same footing,  we  confine the  spectrum of the battery-Hamiltonian in $[-1,1]$ by normalizing it  \cite{ASD} as
\(\frac{[2H_{B}-(e_{max}+e_{min}) \mathbb{I}]}{e_{max}-e_{min}}\),
where $e_{max}$ and $e_{min}$ are the maximum and minimum eigenvalues of the  original Hamiltonian respectively.
We now argue that if the battery is initially prepared as the ground (thermal) state of  spin-$j$ Hamiltonian, we can increase the power of the battery with the increase of the dimension, $j$ when  the charging is performed by the local field.

\subsubsection*{Transverse spin-$j$ $XY$  model as a quantum battery and its charging process}

Before presenting the results, let us first describe the one-dimensional anisotropic quantum $XY$   spin chain consists of $N$ spin-$j$ particles \cite{Schulz, Goroshkov, spin1XXZ, spin1, piersPRB, XYJMM} whose ground (thermal) state acts as an initial state of the battery. In this paper,  we  consider the model having the nearest neighbor interactions among the spins and with open boundary condition. The spin-$j$  $XY$ battery-Hamiltonian, $H^{j}_B$, reads as
\begin{eqnarray}
H^j_B &=& {\frac{1}{2} h \sum_{k=1}^N  S_{k}^z} \nonumber\\
&+& \frac{1}{4}\sum_{k=1}^{N-1} J_{k}[(1+\gamma)S_{k}^x \otimes S_{k+1}^x+(1-\gamma)S_{k}^y\otimes S_{k+1}^y],
\label{eq_mainHamil}
\end{eqnarray}
where $S_{k}^{i}$  is the spin matrices with $i = x,y,z$ acting on site,  $k$. E.g., for spin-1/2 particles i.e., when $j=1/2$, $S_{k}^{i}$ represents the spin Pauli matrices. Here $J_{k}$ is the site-dependent interaction strength between the particles, $h$ is the external magnetic field, and  $\gamma$ is the anisotropy parameter. Notice that \(\gamma=0\) and \(\gamma =1\) correspond to the spin-$j$ $XX$ and Ising models respectively.  When $J_k =J$, i.e., we remove the site-dependence in the coupling, the model reduces to the ordered system, otherwise, it is disordered.  As mentioned before,  we normalize the spectrum of the Hamiltonian between $-1$ and $1$. The magnetic field  applied  to charge the battery  at each site is given by 

\begin{equation}
H_{c}^j=\frac{\omega}{2}\sum_{k=1}^N S_{k}^x,
\label{eq_chargingHamil}
\end{equation}

where $\omega$ is the strength of the magnetic field.


\subsubsection{Increment in power with the variation of dimensionality: Spin-$\frac{1}{2}$ vs. spin-$1$ chain} 
\label{analytics}

Let us prepare the initial state of the battery as the ground state (i.e., the zero-temperature state) of \(H_B^j\) with \(j=1/2\) or \(j=1\) consisting of two spins.    The corresponding \(H_c^j\) is applied to store the energy in the battery. 
We will now show that when interaction strength between the two sites is weak, i.e., when $J_k/|h| =J/|h| = \lambda$ is small, maximum achievable power ($P_{\max}$) increases with the increase of the dimension of the spins. Later we will provide evidence  that the results hold even for a spin chain with four sites containing  higher number of spins. 

\textbf{Proposition 1.} If the initial state of the battery is the ground state i.e., the zero temperature state of the spin-$1$ transverse $XX$-model, the maximum average power of the battery, $P_{\max}$, is higher  than that of the $XX$ model consisting of spin-1/2 particles provided the interaction strength between the sites is weak.

\begin{figure*}
	\includegraphics[width=0.9\textwidth]{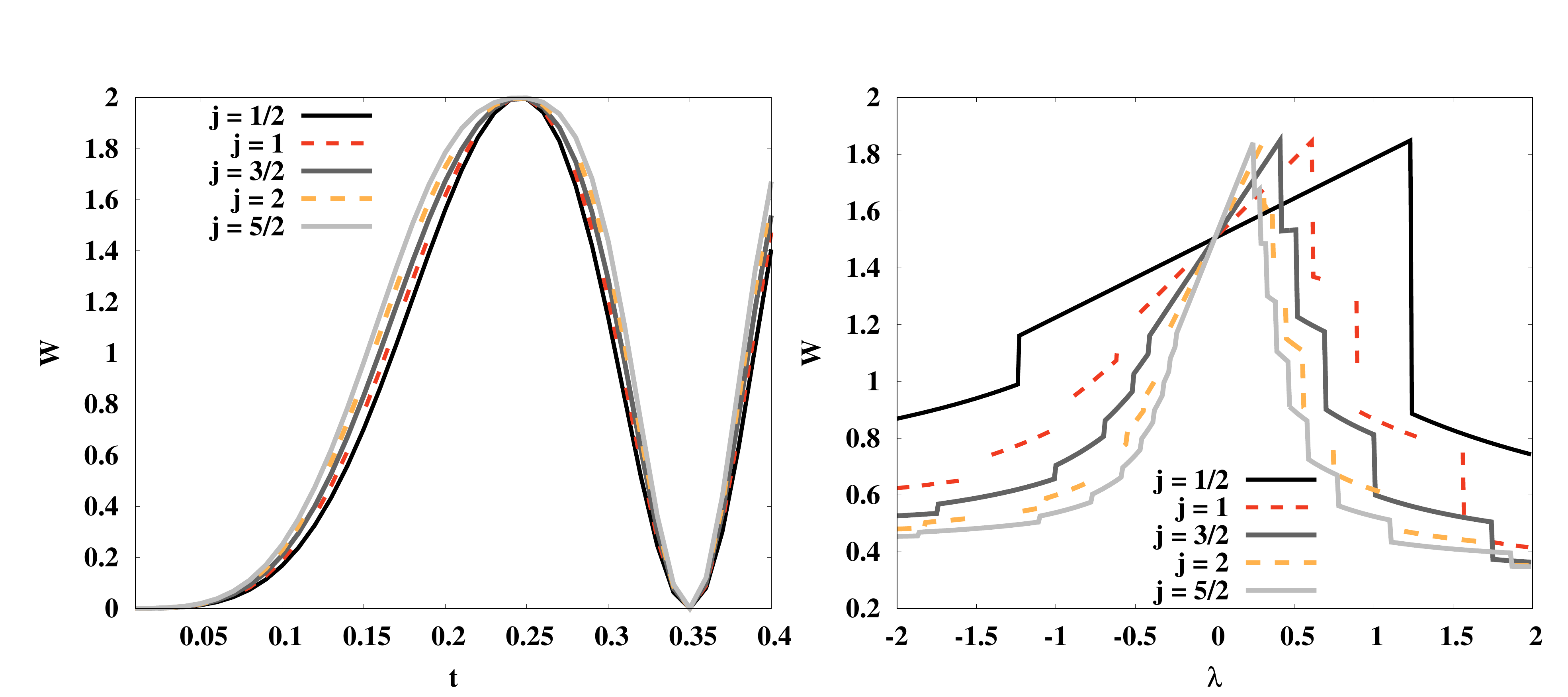}
	\caption{(Color Online.)  Amount of stored work $W (t)$ (ordinate) against time $t$ (abscissa) (left panel) and with $\lambda$ (abscissa) (right panel) for the transverse $XX$ model, i.e., for $\gamma = 0$ with different values of spin-quantum number, $j$. Solid and dashed lines represent  half-integer and integer spins respectively.  Dark, grey and light grey indicate the increment in the dimension in both the cases. For illustration, we fix $\lambda = 0.2$ for the left panel while $t = 2.1$ for the right panel. Here $N=4$. We also notice that in this model,  the extractable work from the battery, the ergotropy \cite{Alicki} coincides with the work output (cf. \cite{PoliniPRL19, quach2020}).  All the axes are dimensionless. }
	\label{fig:wvsJvst}
\end{figure*}

\begin{proof}

Let us first calculate the maximum average power of the battery consists of two spin-$\frac{1}{2}$ particles with  $\gamma = 0$.  Without of loss of generality, we consider the strength of the charging field ($\omega$) to be unity. The Hamiltonian in this case takes the form, 
\begin{equation}
\label{eq_H}
H^{j=1/2}_{B} = \begin{pmatrix}
1 & 0 & 0 & 0\\
0 & 0 & \lambda/2 & 0\\
0 & \lambda/2 & 0 & 0\\
0 & 0 & 0 & -1\\
\end{pmatrix},
\end{equation}
 which has the normalized ground state, 
 $\rho^{j =1/2}(0) = |11 \rangle \langle 11|$ as the initial state. 
The corresponding initial energy is   $W_{initial} = -1$. After the evolution according to \(H_c^{j=1/2}\), the evolved state at time $t$ looks like
\begin{equation}
\label{rho_t_1/2}
\rho^{j=1/2}(t) = \begin{pmatrix}
\frac{B^2}{4} & \frac{ - i B\sin t}{4} & \frac{ - i B \sin t}{4} & -\frac{AB}{4}\\
\frac{-B\sin t}{4i} & \frac{\sin^2 t}{4} & \frac{\sin^2 t}{4} & \frac{A \sin t}{4i}\\
\frac{-B\sin t}{4i} & \frac{\sin^2 t}{4} & \frac{\sin^2 t}{4} & \frac{A \sin t}{4i}\\
-\frac{AB}{4} & \frac{i A \sin t}{4} & \frac{i A \sin t}{4} & \frac{A^2}{4}\\
\end{pmatrix},
\end{equation}
 where $(1 + \cos t) = A$ and, $(1 - \cos t) = B$ and the final energy reads as
\begin{equation}
\label{eq:Wfhalf}
W_{final}= (a \cos t + b \lambda \\sin^{2} t),
\end{equation}
where $a= -0.999698 \approx -1$ (which is actually equal with $W_{initial}$ at $t=0$) and $b=0.25$ with \(\omega =1\) (see Fig. \ref{fig:wvsJvst} with \(N=4\)). Note that \(a\) and \(b\)  depends on \(\omega\). For any values of  \(\lambda\), we can maximize the work with respect to time, thereby obtaining the power. E.g., \(\lambda =0.2\),  we find that 
\(P^{j=\frac{1}{2}}_{max} = \max_{t} \frac {W_{final}-W_{initial}}{t} = 0.7370\) which occurs at time, \(t=2.23\).  
We will restrict  only to the situation when \(\lambda\) is small and positive.  Note that the choice of \(\lambda\) up to which the battery-Hamiltonian performs better than the classical one  depends on \(j\) (see Table \ref{table:jvsJ}) (Notice that when the battery-Hamiltonian in Eq. (\ref{eq_mainHamil})  is without the interaction term, i.e., \(J_k=0\), we refer the model as the classical model of the battery.)

Following the same prescription,  we now evaluate \(P_{\max}^{j=1}\)  for spin-$1$ $XX$ model.  
In the computational basis, the Hamiltonian reads as
\begin{eqnarray}
&& H^{j=1}_{B}  = (|00\rangle \langle 00|-|22\rangle\langle22|)  + \frac{1}{2}(|01\rangle\langle01|+|10\rangle\langle10|) \nonumber \\
&&+ \frac{1}{2}(|12\rangle\langle12|-|21\rangle\langle21|) \nonumber \\ 
 &&+ \frac{\lambda}{2}(|01\rangle\langle10|+|02\rangle\langle11|+|10\rangle\langle20|+|11\rangle\langle21|) + h.c.,
\end{eqnarray}
and the initial state $\rho^{j=1}(0)$ becomes $|22\rangle \langle 22|$. After evolving the state, the final state at time $t$ reads as\\
$|\psi^{j=1}(t)\rangle = a(t)(|00\rangle + |22\rangle) + b(t)|01\rangle + c(t)(|02\rangle + |10\rangle + |20\rangle + |21\rangle) + d(t)|11\rangle + e(t)|12\rangle$.
Here $a(t) = 0.37 + 0.06 e^{-2it} + 0.06 e^{2it} - 0.5 \cos t$, $b(t) = i(0.35 \sin t - 0.16 \sin 2t)$, $c(t) = -0.125 + 0.06 e^{-2it} + 0.06 e^{2it}$, 
$d(t) = -0.25 + 0.13 e^{-2it} + 0.13 e^{2it}$, and
$e(t) = i(-0.35 \sin t - 0.16 \sin 2t)$.\\
In this case,  $W_{initial} = -1$ and $W_{final} = -\cos t + 0.25 \lambda (1 - \cos t)$,  we calculate the power at time t 
which can be represented as 
\begin{eqnarray}
\label{eq:pmax1}
    &&P^{j=1}(t) = a' - b' \cos t+ \cos t( - b' - c' \cos t)\nonumber \\
    &&+ [-2( 1 - a')-c' \cos t + (1 - a')\cos 2t]\cos 2t \nonumber \\
    &&+ 0.05 \sin^{2} t - \frac{\sin t \sin^{2} t}{4} + (1-a') \sin^{2} t,
     \end{eqnarray}
 where \(a', b', \) and \(c'\) are functions of \(\lambda\) as well as \(\omega\).
  By maximizing over time and considering the same \(\lambda =0.2\) as in case of spin-$\frac{1}{2}$ systems,  we obtain $P_{max}^{j=1} = 0.752296$ which is clearly higher than that obtained for spin-$\frac{1}{2}$ systems, i.e.,  $P_{\max}^{j=\frac{1}{2}} < P_{\max}^{j=1} $ with \(\lambda=0.2\). By using Eqs. (\ref{eq:Wfhalf}) and (\ref{eq:pmax1}), we can always find that when \(\lambda\) is weak, the storage capacity of the battery increases with \(j\). For example, with \(\lambda= 0.1\), $P_{\max}^{j=\frac{1}{2}} = 0.7304 < P_{\max}^{j=1} = 0.7371 $ while \(\lambda =0.5\) gives $P_{\max}^{j=\frac{1}{2}} = 0.7608 < P_{\max}^{j=1} = 0.8145 $. 
  
\end{proof}

\textbf{Remark 1.} Although the hierarchy among power in Proposition 1 is proven by comparing  \(j=1/2\) and \(j=1\), it can be shown to be true for other higher dimensional systems as well, for  small values of $\lambda$ (see Fig. \ref{fig:Jvspmaxgamma0.0} and TABLE: \ref{table:jvsJ}).  Such a result can be intuitively understood as follows:  for a weak or negligible  interaction strength,  the spins are initially pointing towards the $z$-direction due to the external magnetic field of the battery-Hamiltonian at zero temperature for a  given dimension. Since the charging field is  applied in the $x$-direction, aligning the spins along the direction of the charging field requires more energy  for driving the system out of equilibrium than that of the battery-Hamiltonian with higher interaction strength which leads to a generation of a higher amount of work in the former case. In addition,   the gap between the maximum and the minimum energy increases with the increase of dimension and hence  the charging field needs to do more work to drive the system away from equilibrium in the higher-dimensional battery, thereby producing a higher amount of power for a fixed value of $\lambda$ and \(\gamma\) compared to a low-dimensional battery. As shown in Fig. \ref{fig:wvsJvst}, the work is oscillating with time, although its   maximum value is fixed. Since \(W(t)\) is a strictly increasing function of time between the initial time and the time when it reaches its maximum value which leads to the power generation, the argument for the stored energy  can also be applied for the power output and hence the similar dimensional advantage in case of power is also observed as depicted in Fig. \ref{fig:Jvspmax}.
 We will show in the next subsection that the results also hold for a moderate value of interaction strength.

\textbf{Remark 2.} Proposition 1 also holds  for the  \(XY\)-spin models, i.e. for spin models with nonvanishing anisotropy parameter, \(\gamma\). For a given dimension, there exists a critical anisotropy parameter above which no ``quantum advantage" can be seen  (see Figs. \ref{fig:Jvspmax} and \ref{fig:jvspmaxforgamma}).  
Since \(\lambda=0\)  in Eq. (\ref{eq_mainHamil}) leads to the battery-Hamiltonian which  is local having no interaction term, the ground or the thermal state of the battery  cannot have any quantum feature like entanglement and hence \(P_{\max}\) which is more than that obtained via \(\lambda=0\) can be termed as quantum advantage.

\subsection{Power of a quantum battery built-up via quantum $XY$ model in arbitrary dimensions}
\label{neumerics}

\begin{figure}
    \includegraphics[height=5.0cm,width=8.0cm]{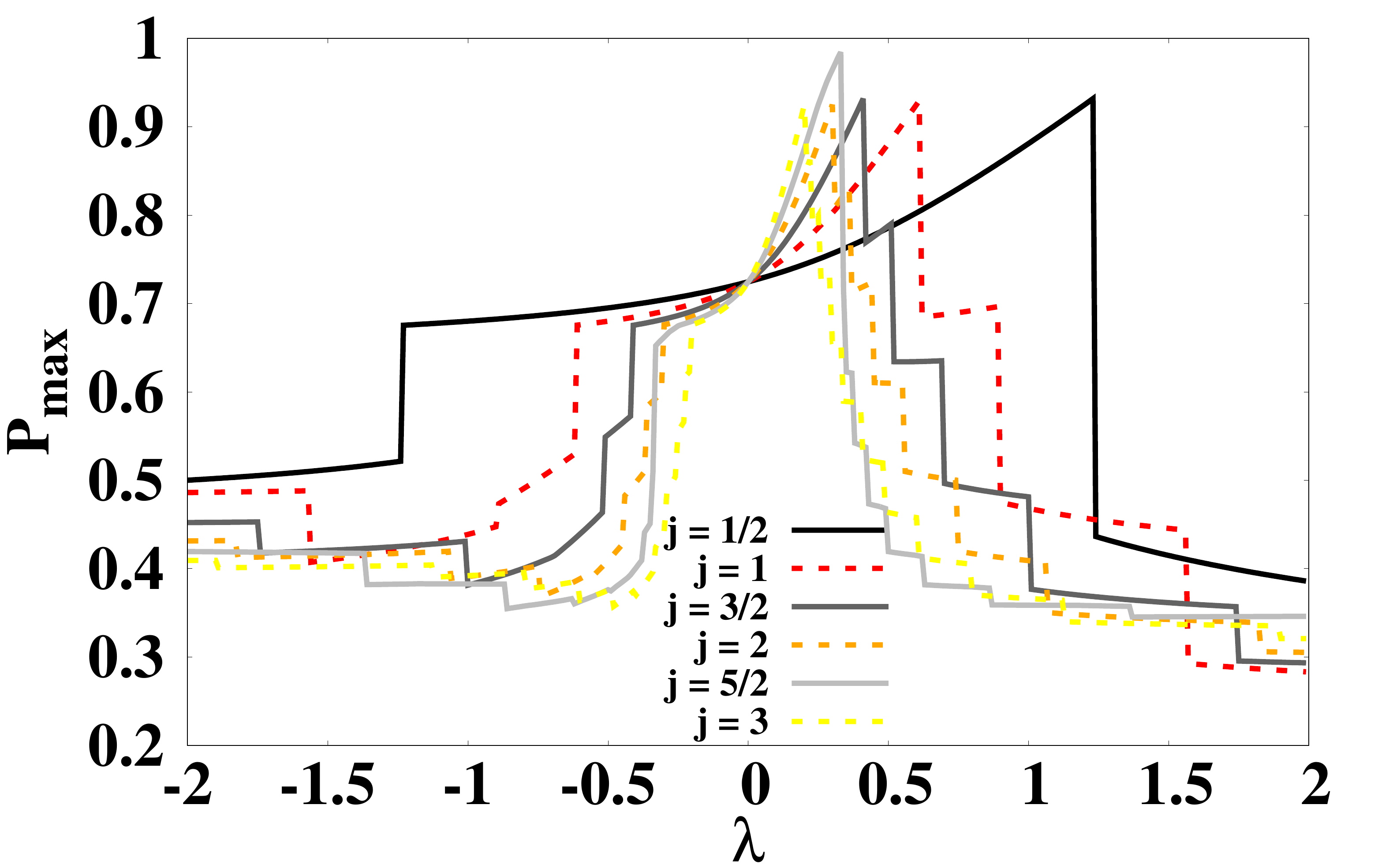}
    \caption{(Color Online.) $P_{max}$ vs. $\lambda$ for the transverse $XX$ model. 
   Solid lines indicate half-integer spins while the dashed ones depict integer spins. Moreover, the dimension of the system increases from dark to light shade in each scenario. In legend, $j$ indicates the spin quantum number of each site.  Here $N=4$. Clearly, there is a region in $\lambda$ where higher dimensional systems perform better than that of the low dimensional models. Both the axes are dimensionless.}
    \label{fig:Jvspmaxgamma0.0}
    
\end{figure}

\begin{table}
\begin{center}
 \begin{tabular}{|c  | c  | c | c | c | c | c |} 
 \hline
 $j$ &  $\frac{1}{2}$ & $1$ & $\frac{3}{2}$ & $2$ & $\frac{5}{2}$ & $3$  \\ [1ex] 
 \hline
 $\lambda_{max} (\gamma = 0) $ &  $1.23$  &  $0.61$ &$0.41$ & $0.30$ &  $0.33$  & $0.20$  \\  [1ex] 
 \hline
 $ \lambda_{max} (\gamma = 0.2)$  & $1.15$  & $0.51$& $0.33$ & $0.25$ & $0.26$& $0.17$ \\ [1ex] 
 \hline
 
 \end{tabular}
\end{center}
 \caption{Maximum values of interaction strength, $\lambda_{max}$,  for which \(P_{\max}\) reaches its maximum for a fixed dimensions of spins, $j$. We choose two anisotropy parameters of the $XY$ model for analysis, specifically,  \(\gamma =0\) and \(\gamma =0.2\). }
\label{table:jvsJ}
\end{table}

By comparing maximum power output  of  the battery prepared by using quantum spin-$\frac{1}{2}$ and spin-$1$  chain, we have already got indication that the storage capacity of the energy in a battery can be enhanced by  increasing the dimension of the model. To  analyse the effect  of  dimensions on work-extraction of the battery, the initial state of the battery is prepared as the  ground state of the anisotropic spin-$j$ $XY$  chain. Let us start  by studying the variation of stored energy, $W(t)$, of the battery with respect to time for a fixed value of interaction strength, and  the behavior of work output against the inetraction strength, $\lambda$ for a fixed value of time (see  Fig.\ref{fig:wvsJvst}).  Here one must note that there is no optimization over time.  Moreover, we notice that \(W(t)\) coincides with the maximum extractable work from the battery, i.e., ergotropy \cite{Alicki}. We observe that both in the oscillatory dynamics of extracted work and in the variation of work with respect to $\lambda$, the dimensional advantage persists, i.e., by increasing the value of spin-quantum number $j$, it is possible to store more energy in the battery. 
To make the analysis more concrete, we compute the maximum power output from the battery and in the rest of the paper, we discuss the performance of the quantum battery in terms of $P_{max}$. \\

\begin{figure}
    \includegraphics[height=5.0cm,width=8.0cm]{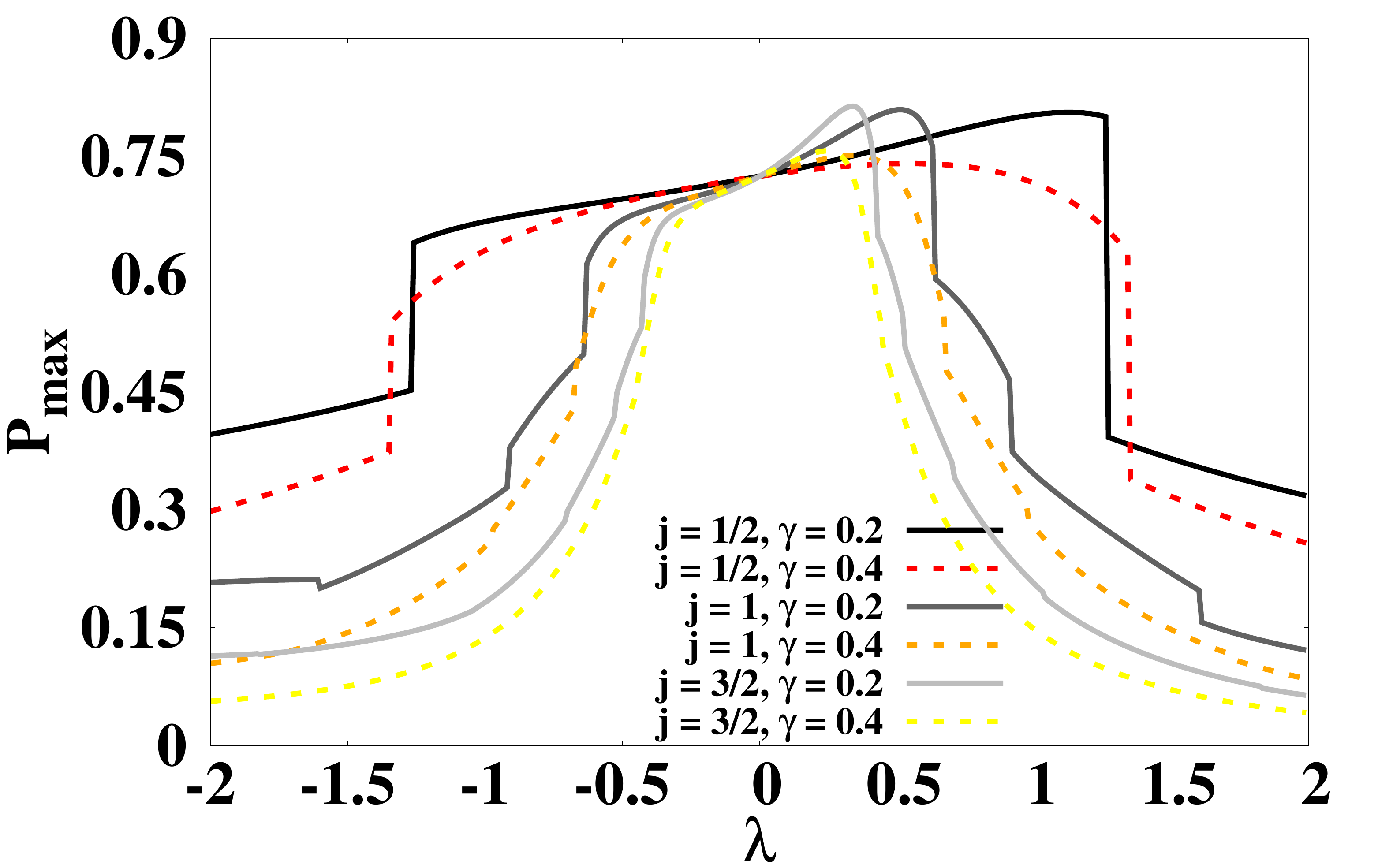}
    \caption{(Color Online.) $P_{max}$ against $\lambda$ of the transverse $XY$ model for different values of $j$ and anisotropy parameters $\gamma$. Solid lines  and dashed lines are for $\gamma =0.2$ and  for $\gamma = 0.4$ respectively.  The value of dimension increases from dark to light grey shades in both the cases. It shows that along with \(\lambda\), anisotropy parameters also play a crucial role to obtain dimensional advantage. Other specifications are same as Fig. \ref{fig:Jvspmaxgamma0.0}. Both the axes are dimensionless. }
    \label{fig:Jvspmax}
\end{figure}

\begin{figure}
    \includegraphics[height=5.0cm,width=8.0cm]{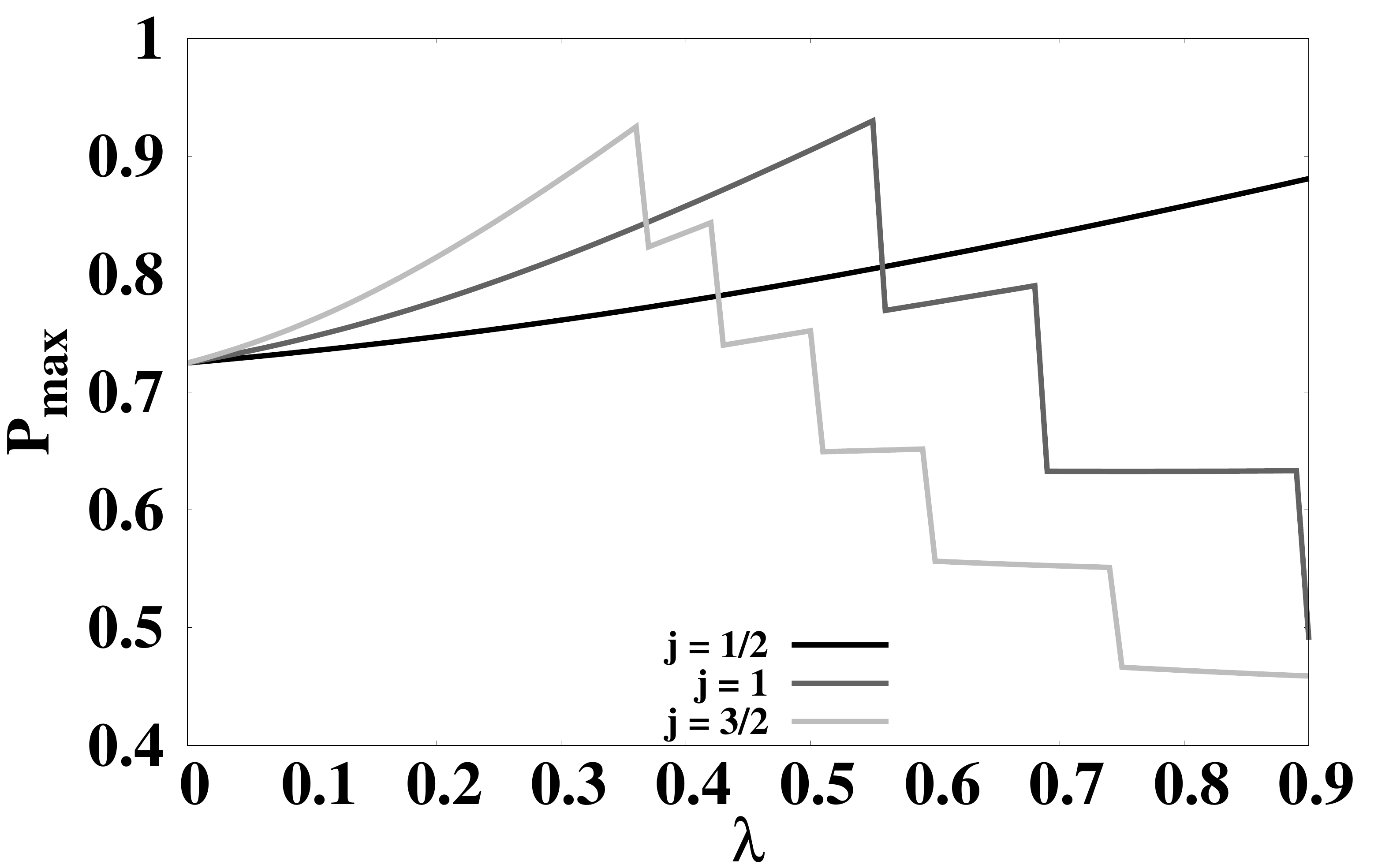}
    \caption{(Color Online.) $P_{max}$ against $\lambda$ of the transverse $XY$ model for different values of $j$ and $\gamma = 0$. Here $N=6$. Both the axes are dimensionless. }
    \label{fig:n6}
\end{figure}

 The entire analysis is performed by considering a spin chain consisting of four sites. We cannot go beyond that number since with $j$, the size of the matrices involved in the computation increases, thereby restricting the numerical simulation of the dynamics.   Note, however,  that the performance remains qualitatively similar even if we consider the battery-Hamiltonian with a large number of spins which we check for small dimensions (see Fig. \ref{fig:n6} for \(N=6\)). The observations regarding the interplay between  the dimension, the interaction strength, and the anisotropy are listed below.



\begin{figure}
    \includegraphics[height=5.0cm,width=8.0cm]{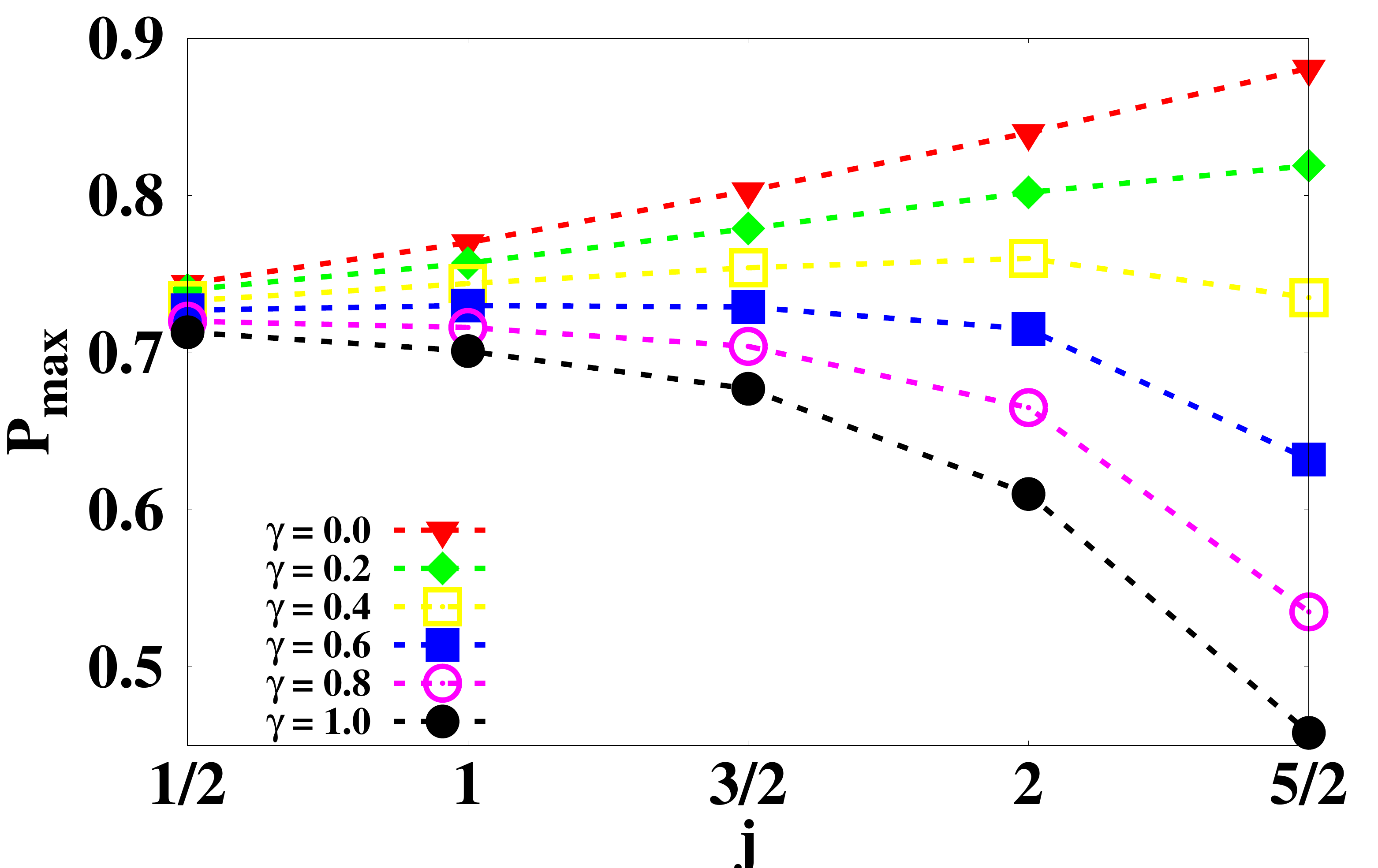}
    \caption{(Color Online.) Variation of $P_{max}$  (ordinate) of the $XY$ battery-Hamiltonian with respect to  spin  quantum numbers, $j$, (abscissa). Different symbols correspond to different strengths of the anisotropy, \(\gamma\). The interaction strength is fixed to $\lambda = 0.2$.  Both the axes are dimensionless.}
    \label{fig:jvspmaxforgamma}
    
\end{figure}

\begin{enumerate}
\item \emph{Trade-off between dimension and interaction strength.}
Let us first carry out the analysis of a quantum battery by fixing $\gamma = 0.0$. As depicted in Fig. \ref{fig:Jvspmaxgamma0.0},  with the increase of dimension of the spins, the maximum average power ($P_{max}$) of the system increases monotonically when $0< \lambda \leq \lambda_{max},$ where \(\lambda_{\max}\) denotes the maximum value of the interaction strength for which \(P_{\max}\) reaches its maximum value. Interestingly, notice that after an  increase for low $j$, \(P_{\max}\) saturates along with the range of \(\lambda\), giving an  advantage, for high values of $j$. In particular, from the figure, we observe that the slope of \(P_{\max}\) remains almost the same with the further increase of \(j\). 
Hence we can possibly conjecture  that in arbitrary dimension, the power output gives a nonclassical enhancement in power in presence of weak interactions.
 As found in Table \ref{table:jvsJ}, \(\lambda_{\max}\) decreases with \(j\). It implies that  the advantage obtained in a higher dimensional spin chain comes at the cost of a more control on the interaction strength. 

\item \emph{Role of anisotropy in power. } Along with the coupling constant, the anisotropy between the interaction strength in the $xy$-plane also plays a crucial role in power extraction. In particular, increase of anisotropy  decreases the enhancement. In other words, the work-output in the battery is more pronounced for the $XX$ model compared to the $XY$ model with \(\gamma \neq 0\) as shown in Fig. \ref{fig:Jvspmax}. 
Since the charging field is in the \(x\)-direction, for non-zero value of \(\gamma\) and a for a fixed dimension, it will be easier for the charging Hamiltonian to
drive the system out of equilibrium, resulting to a lower amount of power generation. 
To show it more precisely, we define a quantity $\gamma_{critical}^{\lambda}$.  For a fixed value of $\lambda$ and $j$, it is defined as the value of the anisotropy parameter  upto which we can get dimensional advantage in terms of power extraction from the quantum battery. Upto the numerical accuracy of four decimal places, we report that $\gamma_{critical}^{0.01} = 0.98$ while $\gamma_{critical}^{0.1} = 0.8$ for all values of $j\leq 3/2$. Moreover, we fix the interaction strength, say, \(\lambda =0.2\) and see the behavior of \(P_{\max}\) with \(j\) for a different anisotropy. 
 Fig. \ref{fig:jvspmaxforgamma} imitates the aforementioned points in a much clarified way with respect to the dimension of the spins.  It is clear from this analysis, that  with the decrease of $\lambda$,  the range of \(\gamma\), i.e.,  $\gamma_{critical}^{\lambda}$  increases for a given value of \(j\). 
\end{enumerate}

\section{Phase dependence of power-output  in spin-$j$ Bilinear-Biquadratic chain   }
\label{haldane}


Upto now, dimensional improvements in the performance of the battery are illustrated by considering quantum $XY$ model. A natural question is to find out whether the enhancement persists even for other one-dimensional battery-Hamiltonian.  To address this question,   the ground state of  the spin-$j$ bilinear-biquadratic Hamiltonian is considered as the initial state of the battery
\cite{Lai74, Takhta82, Suth75, Babu82, Fath91, Fath93, Buchta05}, given by 
\begin{eqnarray}
H_{B}^{j}(\phi) & = & \sum_{k=1}^{N-1} J_{k}[\cos \phi (\overrightarrow {S}_{k} . \overrightarrow { S}_{k+1})\\& +& \sin \phi (\overrightarrow {S}_{k}. \overrightarrow {S}_{k+1})^{2}] +  \frac{h}{2} \sum_{k=1}^{N} S_k^z. \nonumber
\label{BBH}
\end{eqnarray}
Here, $J_{k} \cos \phi$ and $J_{k} \sin \phi$ are the site-dependent interaction strengths, $\overrightarrow{S}_{k}$ are the spin vector acting on the $k$-th site, and $\phi$ is the parameter depending on which a rich phase diagram emerges.
Without the magnetic field, i.e., when \(h=0\),   and for spin-$1$ chain, the Haldane phase appears when \(- \pi/4 < \phi < \pi/4\) while gapless critical phase was found when \(\pi/4 \leq \phi \leq \pi/2\) and the system shows ferromagnetism with \(-3 \pi/4 < \phi <\pi/2\). These three phases and their corresponding phase boundaries are well established although there is a controversy about the critical points of the dimerized phase which is sandwiched between the ferromagnetic and Haldane phases. 

To charge the system,  we consider quadratic charging field $H_{c}^{j}$ of the form,
\begin{equation}
H_{c}^{j} = \omega \sum_{k=1}^{N} \frac{S_{k}^{x}}{2} + \frac{(S_{k}^{x})^{2}}{4},
\label{charging_bbh}
\end{equation} 
where $\omega$ is the strength of the charging field.
\begin{figure}
    \includegraphics[height=5.0cm,width=8.0cm]{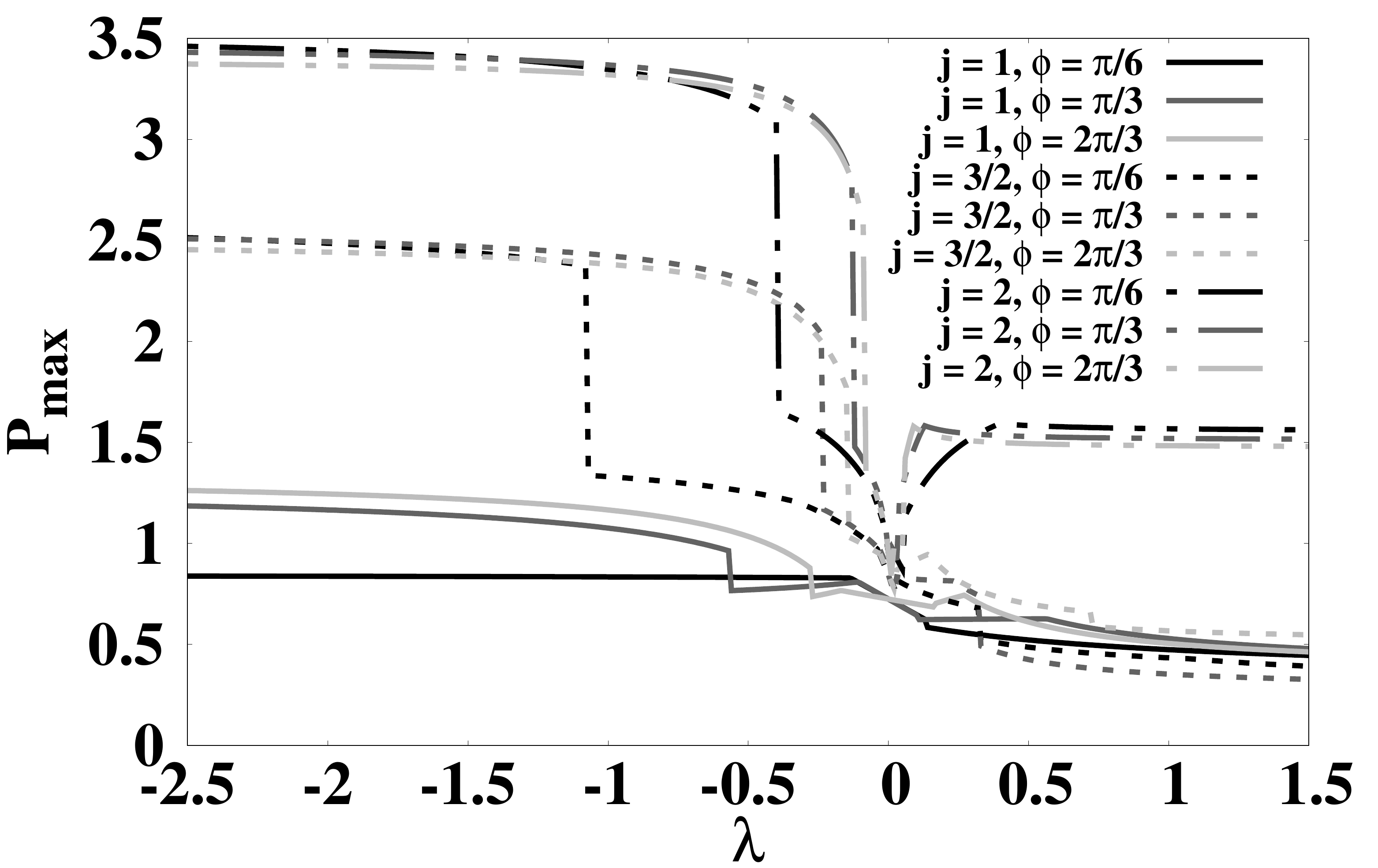}
    \caption{(Color Online.) $P_{max}$ of the BBH model (vertical axis) against $\lambda$ (horizontal axis). Solid, dashed and dot-dashed lines are for the model with $j = 1, 3/2$ and $2$ respectively while black, grey and light grey represent three different values of $\phi =\pi/6$ (Haldane), $\pi/3$ (critical) and $2\pi/3$ (ferromagnetic) respectively, indicating three different phases of BBH model. Both the axes are dimensionless.}
    \label{fig:Jvspmaxforthita}
\end{figure}

To demonstrate the influence of \(\phi\) on the energy storage capacity of a quantum battery when the local magnetic field, given in Eq. (\ref{charging_bbh}) is applied to charge the battery,    we choose three values of $\phi$ -- one is from the Haldane phase, e.g., $\phi=\frac{\pi}{6}$,  the other two  are chosen respectively  from the  critical phase, e.g., $\phi = \frac{\pi}{3}$ and from the ferromagnetic phase, say \(\phi = \frac{2 \pi}{3} \). Unlike the $XY$ model, in all the phases,  we notice striking changes in $P_{max}$ with respect to the interaction strength  and with higher dimension.  The investigations are carried out with the choices of  $j = 1, \frac{3}{2}$ and $2$ to show that 
the power has a  significant dependence on phase $\phi$.

\begin{enumerate}

\item \emph{Phase-dependent dimensional advantage.}

	When \(\lambda  >0\), the interaction does not give any quantum advantage for low dimensional system (cf. \(P_{\max}\) of the battery with \(j=2\))
	which is in stark contrast with the  battery based on the $XY$ spin model. However, the regime, in the parameter space where \(\lambda <0\),  develops some interesting features in the bilinear-biquadratic model and so now onwards, all the analysis are performed for \(\lambda <0\). 	
	 It is evident  from Fig. \ref{fig:Jvspmaxforthita} that the dimensional advantage persists  irrespective  of the value of the phase parameter in the initial state. 
	Nonetheless for $j=1$, the ferromagnetic phase gives beneficial behavior over Haldane and critical phases in terms of $P_{max}$. However, for higher dimensional systems ($j = 3/2$ or $2$), Haldane phase slowly takes over the other two  although the difference between  \(P_{\max}\)  with \(\phi=\pi/6\) and  that of a battery with \(\phi=\pi/3\) or \(\phi=2 \pi/3\) is very small. \\
	\\	
	 In all these cases, we observe two types of improvements in the performance of the battery --  one is due to the increase of spin quantum number and  other one is for the increase of the  interaction strength in the negative direction (see Fig. \ref{fig:Jvspmaxforthita}). Like the $XY$ model, we also notice  that the increment obtained for power via spin-$1$ and spin-$\frac{3}{2}$ is much bigger than that of spin-$\frac{3}{2}$ and spin-$2$ chain. Hence, it can be argued that for a fixed \(\phi\), \(P_{\max}\) saturates to a finite value even in arbitrary large dimension, thereby exhibiting quantum gain in the battery over its classical counterparts.

\end{enumerate}

\begin{figure}
    \includegraphics[height=5.0cm,width=8.0cm]{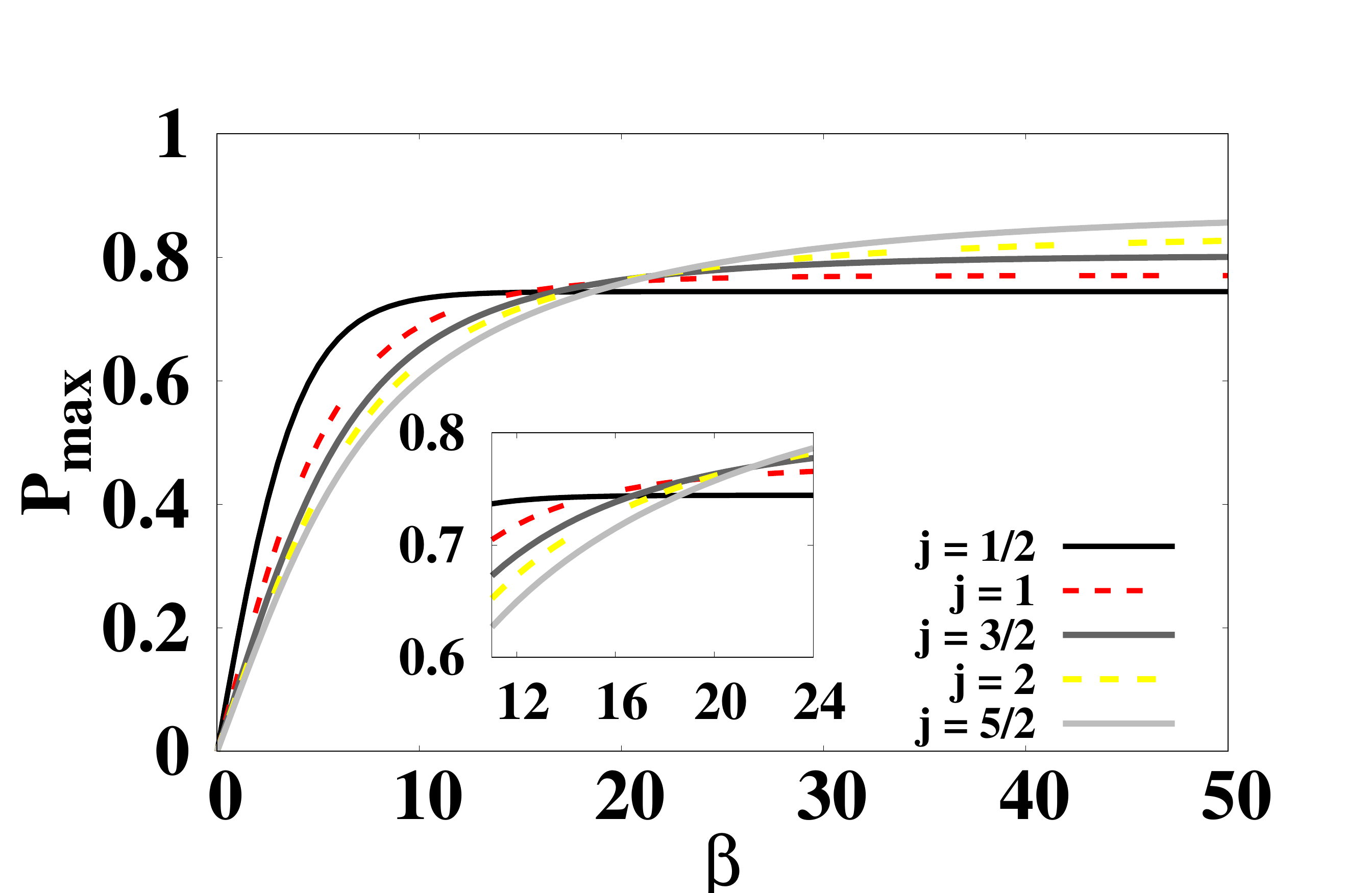}
    \caption{(Color Online.) Dependence of $P_{max}$ (vertical axis) on inverse temperature $\beta$ (horizontal axis) for different values of $j$. Here we take $\gamma = 0.0$ and $\lambda = 0.2$. (Inset) It zooms the region of \(\beta\), where all the lines  cross. The rest of the specifications are same as in Fig. \ref{fig:Jvspmaxgamma0.0}.  Both the axes are dimensionless.}
    \label{fig:betavspmaxforgamma0}
\end{figure}

\begin{figure}
    \includegraphics[height=4.5cm,width=8.5cm]{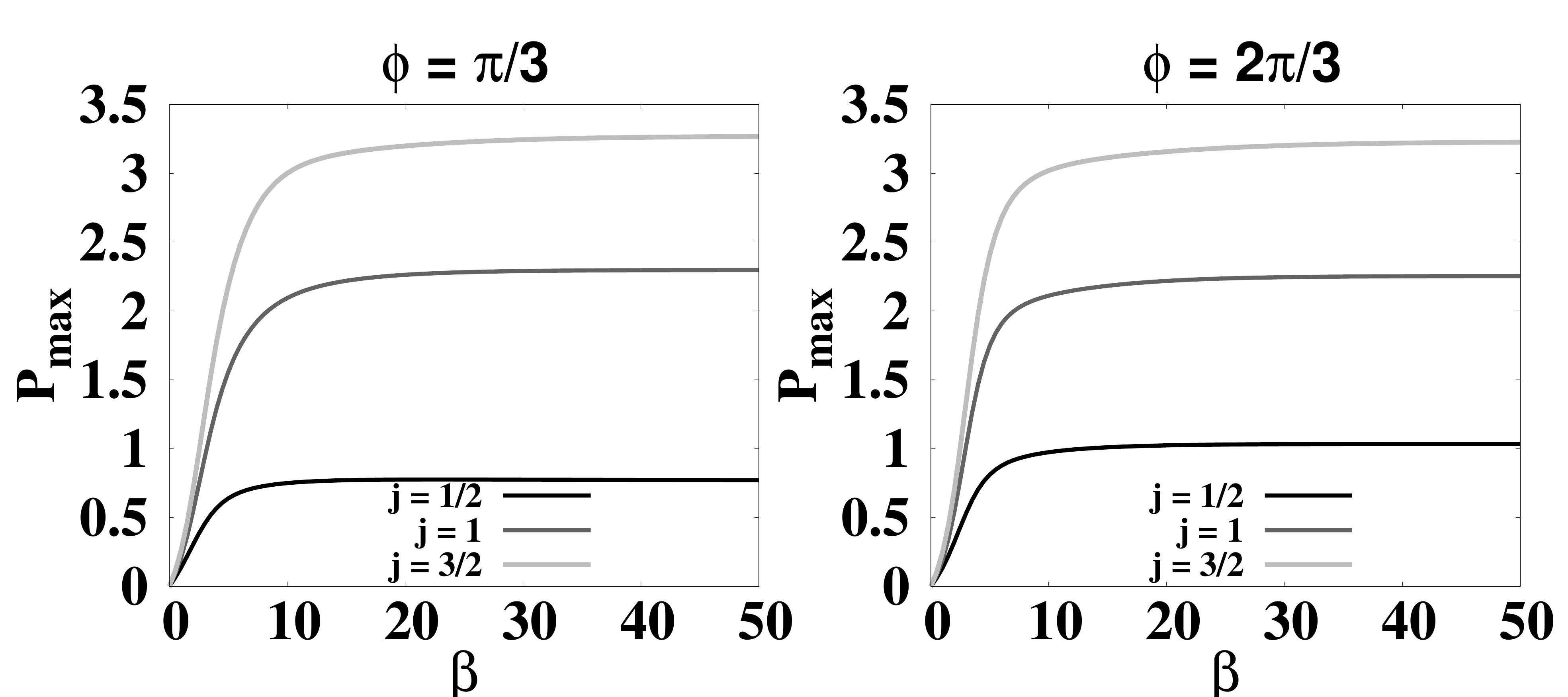}
    \caption{(Color Online.)  $P_{max}$ (ordinate) with  $\beta$ (horizontal axis) for the BBH model. 
    From below, the dimension increases, i.e., from below, systems with $j = 1, 3/2$ and $2$ are plotted. (Left panel) $\phi = \pi/3$  and (right panel) $\phi = 2\pi/3$.  $\lambda = - 0.5$. Both the axes are dimensionless.}
    \label{fig:BBHdis_beta}
\end{figure}

\section{Effects of large spin persists even in presence of Imperfections}
\label{sec:imperfect}

We will now show that the dimensional benefit can also be exhibited  when there is imperfections in the battery. In the preceding section, we always prepare  the initial  state  at zero-temperature. Let us see the consequence on the power if the initial state of the battery is the canonical equilibrium state with a finite temperature. Moreover, we deal with the scenario when the  battery-Hamiltonian is disordered, i.e., the interaction strength is site-dependent. In both the situations, our aim is to determine whether arbitrary large spin helps in the performance of the battery or not. 

\subsection{Temperature-induced power of a quantum battery}
\label{thermal}

We will now probe the situation when the initial state is the canonical equilibrium state, $\rho_{th} = \frac {e^{-\beta' H_{B}^{j}}}{Z}$, where $Z=\mbox{Tr}(e^{-\beta' H_{B}^{j}})$ is the partition function of the system with  inverse temperature $\beta' = \frac{1}{k_{B} T}$ ($k_{B}$ being the Boltzmann constant and $T$ being the absolute temperature). The charging process follows the same unitary evolution governed by the local Hamiltonian, \(H_c^j\) in Eq. (\ref{eq_chargingHamil}). For investigations, we set \(\beta = |h| \beta'\). 

\begin{table}
\begin{center}
 \begin{tabular}{|c  | c  | c | c | c |} 
 \hline
 $j$ & $1$ & $\frac{3}{2}$ & $2$ & $\frac{5}{2}$  \\ [1ex] 
 \hline
 $\beta_{critical} $ &  $15.5$  &  $18.5$ &$20$ & $21$  \\  [1ex] 
 \hline

 \end{tabular}
\end{center}
 \caption{ The cut-off value of the inverse temperature $\beta_{critical}$, of the $XY$ model from which the dimensional advantage is present for a fixed dimensions of spins, $j$. Here,  \(\gamma =0\) and $\lambda = 0.2$. }
\label{table:2}
\end{table}

\emph{Low temperature regime.} For large value of $\beta$, we notice that  higher dimensional spins give a larger amount of power output than that of the low-dimensional systems provided the interaction strength is weak and positive in the $XX$ model as  in Fig. \ref{fig:betavspmaxforgamma0}. It is in a good agreement with the ground state case reported before. To make the analysis more concrete, for the $XY$ model, we compute $\beta_{critical}$, the cut-off value of the inverse temperature above which the dimensional enhancement  occurs, i.e., where  \(P_{\max}^j < P_{\max}^{j + \frac{1}{2}}\). From TABLE: \ref{table:2}, 
we observe that  $\beta_{critical}$ is increasing with the dimension of the spins, $j$ for the $XX$ model. 

On the other hand, for the BBH model, the behavior of $P_{max}$ again turns out to be phase dependent and the trade-off between phases and dimension still exists as mentioned earlier. We observe that higher dimensional battery  remains beneficial, independent of the choice of \(\phi\)s as depicted in Fig. \ref{fig:BBHdis_beta}.


\emph{High temperature region.} Interestingly, however, the opposite picture emerges  for high values of temperature, i.e., with low \(\beta\) for the $XY$ model having a weak coupling constant of the battery-Hamiltonian. Specifically, the spin-chain of low dimension is more beneficial than that of the  model with large spin. Therefore, we observe a critical temperature, $\beta_{critical}$ which separates these two  regions (as shown in the inset of Fig.  \ref{fig:betavspmaxforgamma0}). It is prominent that  with increasing dimension, $\beta_{critical}$ also increases. As argued before, the higher power-generation in the low-dimensional system is possibly due to the fact that  the charging field in this case requires more energy to drive the system away from equilibrium compared to the higher dimensional systems which can  involve more number of eigenstates in the process.\\
 Such a critical temperature  is not observed in case of the BBH model, i.e.,  in the high temperature regime, higher dimensional systems continue to generate a higher amount of power compared to a battery with low dimensions as shown in  Fig. \ref{fig:BBHdis_beta}. 

\begin{figure}
    \includegraphics[height=5.0cm,width=8.0cm]{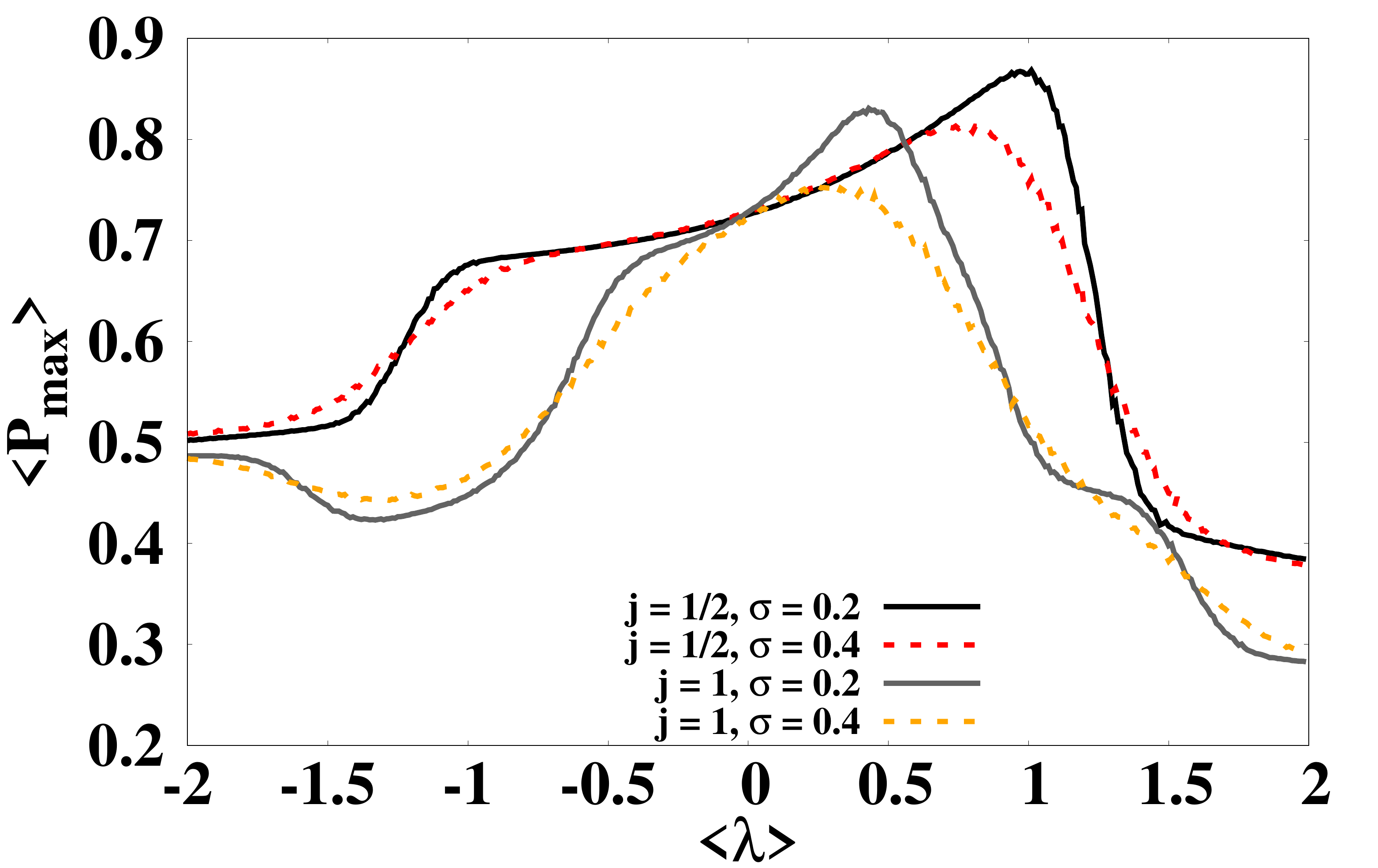}
    \caption{(Color Online.) Quenched averaged  power output,  $\langle P_{max}\rangle$ (ordinate) with respect to $\langle \lambda\rangle$ (abscissa) for the disordered $XX$ model. Solid and dashed lines represent $\sigma_\lambda =0.2$ and $0.4$ respectively while black lines depict spin-$\frac{1}{2}$ systems and light grey lines are for spin-$1$ particles. Both the axes are dimensionless.}
    \label{fig:Jvspmaxforsigma}
\end{figure}

\begin{figure*}
    \includegraphics[width=0.9\textwidth]{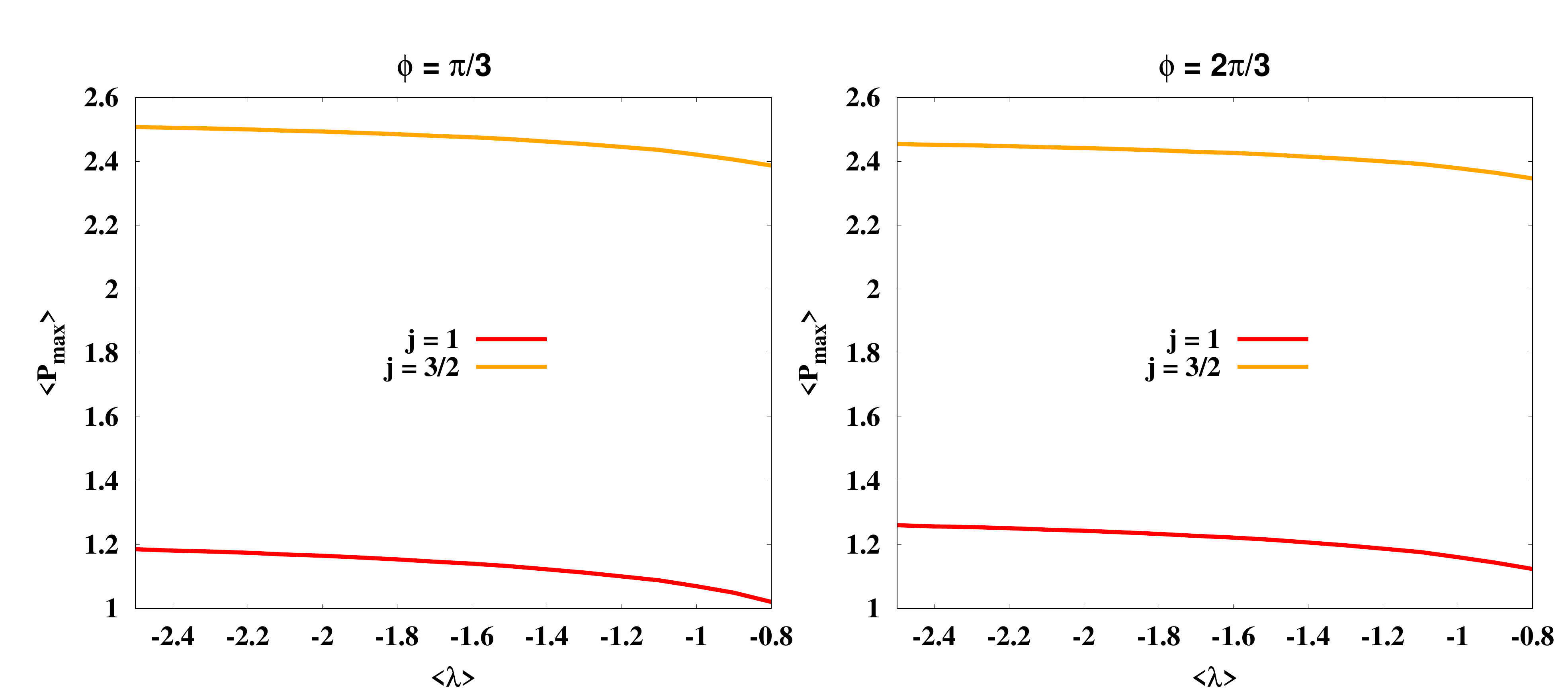}
    \caption{(Color Online.)  $\langle P_{max}\rangle$ (vertical axis) against $\langle \lambda\rangle$ (horizontal axis) for disordered BBH model. Dark and light lines represent  the battery with \(j=1/2\) and \(j=1\) respectively.  Here we have taken \(\sigma_{\lambda} =0.2\). (Left panel)  Critical phase with \(\phi = \frac{\pi}{3}\)  and (right panel) ferromagnetic phase with \(\phi = \frac{2 \pi}{3}\) . Both the axes are dimensionless.}
    \label{fig:BBHdis}
\end{figure*}

\subsection{Impurity along with large spin leads to increment in power }
\label{dis}

We will now concentrate on a battery-Hamiltonian which has some defects occurred due to imperfections in the preparation process or due to environmental influence. Here we assume that the change in presence of disorder is very slow compared to the dynamics of the system and hence we can perform ``quenched averaging'' of the physical quantity. 

\emph{Quench averaging. } In both the models considered in the preceding section,  we choose randomly the site-dependent interaction strength, \(J_k/|h|\) from Gaussian distribution with mean \(\langle J \rangle/|h| \equiv \langle \lambda\rangle\) and standard deviation, \(\sigma_{\lambda}\). We then compute the maximum power for each such choices and  repeat the procedure for several times.   At the end, we perform averaging over all such realizations, to obtain the quenched averaged power,  \(\langle P_{\max}\rangle\). The number of realizations performed depends on the convergence of the physical quantity. In our  study, we perform \(2000\) realizations and observe that 
\(\langle P_{\max}\rangle\) converges upto second decimal points. 

\begin{enumerate}

    \item \emph{Variation of Power for disordered $XX$ model with spin quantum number. }  The transverse disordered $XX$ spin chain shows dimensional improvements for low values of \(\langle \lambda \rangle >0 \) with small disorder strength, $\sigma_{\lambda}$ (see  Fig . \ref{fig:Jvspmaxforsigma}), i.e.,  \(0< \langle \lambda \rangle <\langle \lambda_{\max}\rangle \),  $\langle P_{\max} \rangle^{j+\frac{1}{2}} > \langle P_{\max} \rangle^{j}$. It clearly shows that even if impurity is present in the system, dimensional upgrading in terms of power storage capacity  can  be obtained. 
    
    However, such an advantageous situation vanishes if one increases the strength of the disorder, i.e., the value of $\sigma_{\lambda}$. It is again  due to  the trade-off relation between the interaction strength and the dimensions mentioned in the ordered case. In particular, when interaction strengths are chosen randomly from the Gaussian distribution, we know that most of the times, interaction strengths lie between \(\langle \lambda \rangle \pm 3 \sigma_{\lambda}\) and hence comparing the ordered scenario, we can provide dimensional benefit only for small \(\sigma_{\lambda}\).  Precisely, we find that  when $\sigma_{\lambda}>0.3$,  higher values of $\lambda$ are coming into play, and   for the ordered $XX$ model, we found  (Table \ref{table:jvsJ}) that high values of $\lambda$ do not show any advantage with $j$.  Hence, when  strength  of the disorder is strong, disordered systems cannot demonstrate any large spin-benefit. 
    Moreover,  we notice that there exists a small region with \(\langle \lambda \rangle <0\) where  disordered spin-$1$ $XX$ model gives more power output in comparison with the ordered ones, which is not present in case of spin-$\frac{1}{2}$ particles.
 
    \item \emph{Randomly chosen interaction strength for BBH.}
    As reported before, in the ordered case, when \(\lambda <0\) and  \(\phi = \frac{2 \pi}{3}\), BBH model with $j = 1$  gives a higher \(P_{\max}\) than that of the initial state prepared in a different phase with same spin-quantum number $j$.
  Similar situation remains true even in presence of disorder. To illustrate it, we choose \(\langle \lambda \rangle <0\), and \(\phi = \frac{ \pi}{3}\) as well as \(\phi = \frac{2  \pi}{3}\). Moreover,  we find that with moderately high \(\sigma_{\lambda}\), quenched averaged power, \(\langle P_{\max}\rangle\) of the  spin-$\frac{3}{2}$ BBH model is significantly higher than that of the  spin-$1$ case, as depicted in Fig. \ref{fig:BBHdis}.

 \end{enumerate}

\section{conclusion}
\label{conclusion}
In recent times, the ever-increasing demand for energy and limited resources in a classical world is a great threat to humankind. Technologies based on quantum mechanical principles were shown to overcome the critical situation. In this direction, the classical battery serves a crucial role  by converting chemical energy to an electrical one. Nowadays with the emergence of quantum machinery, it is possible to design \emph{quantum battery}  \cite{Dou_2020, santos2019, tabesh2020, Crescente_2020, PoliniPRL} which is much smaller in size and way more effective in terms of accumulating accessible energy than the classical ones.
From the beginning of its discovery,  it is customary to model a quantum battery consisting of spin-$\frac{1}{2}$ particles. Very recently, three-level product states are considered as the initial state of the battery which is charged via interacting Hamiltonian.  

We here constructed a battery whose initial state is the ground state of an interacting spin-$j$ model and the charging is performed via local unitary operations. 
We showed the beneficial effects of \emph{dimension} on the performance of the quantum battery. In particular, The ground state, as well as thermal state of the transverse $XY$ spin-chain and bilinear-biquadratic Heisenberg (BBH) Hamiltonian with $j$-dimensional spins,  are used to demonstrate  that with the  increase  of dimensions, the power extraction from the system increases.  Specifically, we found that in case of the spin-$j$ $XY$ model, the power-output depends on the interaction strength as well as the anisotropy parameter while the phase of the initial state plays an important role for the BBH-based battery. 

Moreover, in a more realistic situation, we exhibited that the  dimensional benefit is robust against the impurities in the battery-Hamiltonian or at finite temperature. Notice that both of them naturally appear during implementations. Results obtained here are  counter-intuitive in the sense that, by increasing the dimensions, it is believed that we  typically  approach  the classical regime although   advantages persist even with arbitrarily large spin which cannot be obtained via interaction-free battery model.

\section*{acknowledgement}
We acknowledge the useful discussions with Tanoy Kanti Konar and the support from the Interdisciplinary Cyber Physical Systems (ICPS) program of the Department of Science and Technology (DST), India, Grant No. DST/ICPS/QuST/Theme- 1/2019/23. We  acknowledge the use of \href{https://github.com/titaschanda/QIClib}{QIClib} -- a modern C++ library for general purpose quantum information processing and quantum computing (\url{https://titaschanda.github.io/QIClib}) and cluster computing facility at Harish-Chandra Research Institute.

\bibliography{bib}
\bibliographystyle{apsrev4-1}
\end{document}